\documentstyle[12pt]{article}
\def\baselinestretch{1.24}
\begin{document}
\pagestyle{plain}

\bibliographystyle{unsrt}    

\newcommand{\st}{\scriptstyle}
\newcommand{\sst}{\scriptscriptstyle}
\newcommand{\mco}{\multicolumn}
\newcommand{\epp}{\epsilon^{\prime}}
\newcommand{\vep}{\varepsilon}
\newcommand{\ra}{\rightarrow}
\newcommand{\ppg}{\pi^+\pi^-\gamma}
\newcommand{\vp}{{\bf p}}
\newcommand{\ko}{K^0}
\newcommand{\kb}{\bar{K^0}}
\newcommand{\al}{\alpha}
\newcommand{\ab}{\bar{\alpha}}
\def\be{\begin{equation}}
\def\ee{\end{equation}}
\def\bea{\begin{eqnarray}}
\def\eea{\end{eqnarray}}
\def\CPbar{\hbox{{\rm CP}\hskip-1.80em{/}}}

\def\ap#1#2#3   {{\em Ann. Phys. (NY)} {\bf#1} (#2) #3.}
\def\apj#1#2#3  {{\em Astrophys. J.} {\bf#1} (#2) #3.}
\def\apjl#1#2#3 {{\em Astrophys. J. Lett.} {\bf#1} (#2) #3.}
\def\app#1#2#3  {{\em Acta. Phys. Pol.} {\bf#1} (#2) #3.}
\def\ar#1#2#3   {{\em Ann. Rev. Nucl. Part. Sci.} {\bf#1} (#2) #3.}
\def\cpc#1#2#3  {{\em Computer Phys. Comm.} {\bf#1} (#2) #3.}
\def\err#1#2#3  {{\it Erratum} {\bf#1} (#2) #3.}
\def\ib#1#2#3   {{\it ibid.} {\bf#1} (#2) #3.}
\def\jmp#1#2#3  {{\em J. Math. Phys.} {\bf#1} (#2) #3.}
\def\ijmp#1#2#3 {{\em Int. J. Mod. Phys.} {\bf#1} (#2) #3.}
\def\jetp#1#2#3 {{\em JETP Lett.} {\bf#1} (#2) #3.}
\def\jpg#1#2#3  {{\em J. Phys. G.} {\bf#1} (#2) #3.}
\def\mpl#1#2#3  {{\em Mod. Phys. Lett.} {\bf#1} (#2) #3.}
\def\nat#1#2#3  {{\em Nature (London)} {\bf#1} (#2) #3.}
\def\nc#1#2#3   {{\em Nuovo Cim.} {\bf#1} (#2) #3.}
\def\nim#1#2#3  {{\em Nucl. Instr. Meth.} {\bf#1} (#2) #3.}
\def\np#1#2#3   {{\em Nucl. Phys.} {\bf#1} (#2) #3.}
\def\pcps#1#2#3 {{\em Proc. Cam. Phil. Soc.} {\bf#1} (#2) #3.}
\def\pl#1#2#3   {{\em Phys. Lett.} {\bf#1} (#2) #3.}
\def\prep#1#2#3 {{\em Phys. Rep.} {\bf#1} (#2) #3.}
\def\prev#1#2#3 {{\em Phys. Rev.} {\bf#1} (#2) #3.}
\def\prl#1#2#3  {{\em Phys. Rev. Lett.} {\bf#1} (#2) #3.}
\def\prs#1#2#3  {{\em Proc. Roy. Soc.} {\bf#1} (#2) #3.}
\def\ptp#1#2#3  {{\em Prog. Th. Phys.} {\bf#1} (#2) #3.}
\def\ps#1#2#3   {{\em Physica Scripta} {\bf#1} (#2) #3.}
\def\rmp#1#2#3  {{\em Rev. Mod. Phys.} {\bf#1} (#2) #3.}
\def\rpp#1#2#3  {{\em Rep. Prog. Phys.} {\bf#1} (#2) #3.}
\def\sjnp#1#2#3 {{\em Sov. J. Nucl. Phys.} {\bf#1} (#2) #3.}
\def\spj#1#2#3  {{\em Sov. Phys. JETP} {\bf#1} (#2) #3.}
\def\spu#1#2#3  {{\em Sov. Phys.-Usp.} {\bf#1} (#2) #3.}
\def\zp#1#2#3   {{\em Zeit. Phys.} {\bf#1} (#2) #3.}

\setcounter{secnumdepth}{2} 


\begin{titlepage}
\renewcommand{\thefootnote}{\fnsymbol{footnote}}
\begin{center}

{\Large \bf Theoretical Physics Institute\\
University of Minnesota}
\end{center}
\vspace{1cm}
\begin{flushright}
TPI-MINN-95/34-T~\\
UMN-TH-1420-95~\\
hep-ph/9512419~\\
\end{flushright}
\vspace{1cm}
\begin{center} {\Large
{\bf REVIEW OF SELECTED TOPICS IN HQET$^{\dagger}$}}
\footnotetext{$^{\dagger}$ The work is
supported in part by DOE under the grant DE-FG02-94ER40823.}\\
\vspace*{.5cm}
{\Large A. Vainshtein}  \\
\vspace{.4cm}
{\normalsize
 {\it  Theoretical Physics Institute, University of Minnesota,\\
Minneapolis, MN 55455\\{\rm and}\\
Budker Inst. of Nuclear Physics, Novosibirsk, 630090, Russia}}

\vspace{1.5cm}

{\Large{\bf Abstract}}
\end{center}
\vspace{0.4cm}
\centering{\begin{minipage}{5.83truein}

A few topics on the expansion in heavy quark mass are discussed.
The theoretical framework is the Wilson Operator Product
Expansion rather than the Heavy Quark Effective Theory.
\end{minipage}}
\vspace{2cm}

\begin{center}
{\it Invited talk at the European Physical Society\\
Conference on High Energy Physics,\\
Brussels, July 1995;\\
to appear in the Proceedings}
\end{center}
\end{titlepage}

\newpage

\section{Introduction}
The notion of Heavy Quark Effective Theory (HQET) was introduced
by Eichten, Hill\cite{eichten} and Georgi\cite{georgi}. Then
HQET  and its applications were
actively developed by many
authors (see a recent review\cite{neubert}).
The theory is based on the smallness of the parameter
$\Lambda_{QCD}/m_Q$, i.e. the ratio of characteristic hadronic scale to the
scale given by heavy quark mass $m_Q$, and on the existence of
the limit $m_Q \rightarrow \infty$.

The recent development is concentrated mostly on preasymtotic
effects, i.e. on the study of nonpertubative corrections
$(\Lambda_{QCD}/m_Q)^n$. The main theoretical tool is the Operator
Product Expansion (OPE) introduced by Wilson\cite{wilson}
 which allows a separation of
short and large distances.  What is a relation
between HQET and OPE? In a sense HQET is just a
particular application of OPE. However, the way the HQET was implemented
is not entirely consistent with the OPE. One of basic quantities in
HQET is the pole quark mass. The existence of infrared contributions in
the pole quark mass leads to problems when $(1/m_Q)^n$ correction are
taken into account. At this level the standard HQET does not exist as a
quantum field theory\cite{pole,bb}. For this reason I prefer to
refer to the OPE rather to the HQET.

There is a clear analogy between the use of OPE for heavy flavour physics
and classical applications of OPE to $e^+ e^-$ annihilation into
hadrons and to deep inelastic scattering. First, heavy flavour states
can be viewed as ground states of light flavours but in the presence
of almost static gluon field produced by a heavy quark (as different from
vacuum or nucleon states). Second, the analogy between short distance
probes is the analogy between, say, the total cross section of hadron
production in $e^+ e^-$ annihilation versus the total semileptonic
widths of heavy flavours. The heavy quark mass $m_Q$ plays the role
similar to the total energy $W$ in $e^+ e^-$ collisions defining the
scale for perturbative and nonperturbative corrections.

Let me finish this short introduction presenting the partial list of
topics where a theoretical understanding was strongly advanced
during recent years:\\
$\bullet$ Corrections $1/m_Q^2$ to inclusive widths. Spectra near
end-points -- QCD description of the ``Fermi motion'' of heavy quark.\\
$\bullet$ Pole mass: infrared renormalons and $m_Q^{pole}$ uncertainty,
the necessity of normalization point, the correct OPE construction.\\
$\bullet$ Sum rules for heavy flavour transitions. New sum rules,
$(1/m_Q)^n$ correction the known ones, lower bound for the average
kinetic energy of heavy quark.\\
$\bullet$ Extraction of $|V_{cb}|$ from exclusive ($B \rightarrow D^*
l\nu$) and inclusive ($B\rightarrow X_c l\nu$) processes.\\
$\bullet$ Status of semileptonic branching ratio.

\section{Total Widths}
As an example of theoretical predictions let us present a result for
the total width of semileptonic decay $B\rightarrow X_u l\nu$,
\begin{equation}
\Gamma (B\rightarrow X_u l\nu)= \frac{G_F^2 m_b^5 |V_{ub}|^2}{192
\pi^3} \left[ 1 - \frac{\mu_{\pi}^2}{2m_b^2} - \frac{3 \mu_G^2}{2m_b^2}
\right]\;.
\label{gamma}
\end{equation}
Here $\mu_G^2$ and $\mu_{\pi}^2$ are defined as
\be
\mu_G^2=\frac{1}{2m_b} \langle B|{\bar
b}\frac{i}{2} \sigma_{\mu\nu} G^{\mu\nu} b| B\rangle ,\;\;\;\;
\mu_{\pi}^2=\frac{1}{2m_b} \langle B|{\bar
b}(i{\vec D})^2 b |B \rangle.
\label{mug}
\end{equation}
The numerical value of $\mu_G^2$ is known from the hyperfine
splitting of $B^*$ and $B$ mesons:
\begin{equation}
\mu_G^2=\frac{3}{4}(M_{B^*}^2 - M_B^2) \approx 0.36\,{\rm GeV^2}\;.
\label{mugnum}
\end{equation}
Note that for baryons (besides $\Omega_Q$) $\mu_G^2=0$.
As for $\mu_{\pi}^2$ it is not yet extracted from experimental data,
there is only the number which follows from QCD sum rules\cite{ballbr},
$\mu_{\pi}^2=0.5\pm 0.1$.

Numerically power corrections diminish the width determined in equation
\ref{gamma} by
about 3\%. In a similar fashion corrections which are parametrically
$1/m_b^2$ to other
semileptonic and nonleptonic widths are expressed via the same
$\mu_G^2$, $\mu_{\pi}^2$. The absence of corrections of the first
power in $1/m_Q$ is specific for QCD as it was pointed out first
in paper\cite{cgg} (with some reservations about overall normalization).
Explicitly terms $1/m_Q^2$ were calculated in\cite{buvbs} where
the absence of the $1/m_Q$ corrections was stated for the
normalization as well.

Corrections $1/m_Q^2$ are ``spectator blind'', i.e. do not depend
on the flavour of light quark, say, in a heavy meson. The
dependence shows up at the level of $1/m_Q^3$ terms\cite{shifvol}
which numerically are comparable with $1/m_Q^2$ terms.
The overall fit for lifetimes of charm and beauty hadrons looks
satisfactory\cite{bigi} with $\Lambda_b$ as an exception.
Experimentally $\tau (\Lambda_b)/\tau (B_d)=0.76\pm 0.06$ with 0.9
as a preferable theoretical number.

\section{End-point Spectra}
Julia Ricciardi discussed in her talk at this Conference
recent papers\cite{aligreub,dsu} devoted to analysis of photon
spectrum in $B\ra X_s \gamma$ inclusive decay. I would like to
add some comments on the topic.

It was realized long ago that the window
$m_b/2<E_{\gamma}<M_B/2$ in the photon spectra which is empty on
the parton level is filled up only due to the nonperturbative
effect of the heavy quark motion, and phenomenological models
accounting for the effect were suggested\cite{alipietacm}.

What is the QCD (i.e. model independent one) description?
It was worked out in papers\cite{jafferandall}$^-$\cite{fjmw}.
Near end-points the
energy release for the light quark system at the final state is not
of the order of $m_b$ which is much larger than the QCD scale
$\Lambda_{QCD}$ but of the order of ${\bar \Lambda}=M_B-m_b$,
which is $\sim \Lambda_{QCD}$. Thus operators of high dimension
in the OPE are not suppressed by small coefficients ($\propto 1/m_Q^n$)
and the summation is needed. This summation is a natural generalization
of the OPE procedure for deep inelastic processes.

The quantity which substitutes $Q^2$ is $K^2$,
\be
K^2=-k^2=-(m_b v_\mu - q)^2=2m_b(E_{\gamma} -\frac{m_b}{2}),
\label{k2}
\ee
where $q$ is the lepton pair momentum and $v_\mu = p^B_\mu/M_B$
is 4-velocity of $B$ meson.
At the end-point region $K^2\sim m_b {\bar \Lambda}$ which is
much larger than ${\bar \Lambda}^2$ but still much less than $m_b^2$.
Perturbative corrections are governed by $\alpha_s (K^2)$.
Nonperturbative terms are given by the following sum:
\be
\langle B|{\bar b}\frac{2}{K^2} \sum_{n=0}^\infty \left(
\frac{2k\pi} {K^2} \right)^n b|B \rangle ,
\label{exp}
\ee
where $\pi_\mu=iD_\mu$. All terms in this expansion are of the
same order at the end-point region and present moments of
distribution function $F(x)$. The scaling variable $x$ (an analog
of Bjorken $x$) is defined as
$$x=\frac{K^2}{2{\bar \Lambda}(kv)}\approx
(q_0 + |{\vec q}| -m_b)/{\bar \Lambda}\;.$$
{}From  equation \ref{exp} first few moments of $F(x)$ are as follows:
$$
\int dx F(x)=1,\; \int dx x F(x)=0,\; \int dx x^2
F(x)=\frac{\mu_\pi^2} {3 {\bar \Lambda}^2},
$$
\be
\int dxx^3 F(x)=\frac{1}{6{\bar \Lambda}^3} \frac{1}{2M_b}
\langle B|{\bar b}\gamma_0 b\cdot g_s \sum_q {\bar q}\gamma_0 q
|B\rangle.
\label{moments}
\ee
The distribution function $F(x)$ is universal in the sense that it
defines spectra of different processes where final quarks are
relativistic. In particular, it is the case for $B\ra X_s
\gamma$ and $B\ra X_u l {\bar \nu}$ decays,
$$
\frac{d\Gamma(B\ra X_s \gamma)}{dE_\gamma}=\Gamma_0^{s \gamma}\,
\frac{2F(x)}{{\bar \Lambda} } ;
$$
\be
\frac{d\Gamma(B\ra X_u l {\bar \nu})}{dE_l dq^2
dq_0}=\frac{\Gamma_0^{ul\nu}}{m_b^5}\,
\frac{2F(x)}{{\bar \Lambda}}  \frac{24(q_0-E_l)(2m_b E_l
-q^2)}{m_b -q_0}\;.
\label{spectr}
\ee
Then the lepton spectrum near the end point is proportional to
$$
\int_{(2E_l -m_b)/{\bar \Lambda}}^1 dx F(x)\;.
$$
Near the end points perturbative corrections are enhanced and
the summations of leading (double) logs as well as
subleading is required. The resulting perturbative kernel should
be convoluted
with the distributions given in equation \ref{spectr}.
The most advanced realization of this program given in the
paper\cite{dsu}.

Let me emphasize that an absolutely different distribution function
appears when the final quark is slow. The realistic case of $B
\ra X_c l {\bar \nu}$ decay is a mixed one but the $c$ quark is
predominantly nonrelativistic over phase space in this decay.

\section{Irrelevance of Pole Mass and Infrared Renormalons}
Let us start from the one-loop correction to the heavy quark mass $m_Q$.
This correction is described by a simple diagram with Coulomb quanta
exchange along fermionic line. The integration over the gluon
momentum $k$ is limited by $\mu_0$ from above and by $\mu$ from below,
$\mu_0 \ll m_Q$. The result looks as follows:
\be
m_Q(\mu) = m_Q(\mu_0) + \frac{2\pi}{3} \frac{\alpha_s}{\pi}
(\mu_0 - \mu).
\label{m_Q}
\ee
This equation presents running of $m_Q$ in the range of $\mu \ll
m_Q$ and has a simple meaning of accounting for the Coulomb energy
 $2\alpha_s/3 r_0$. The equation \ref{m_Q} reflects the infrared
stability of the mass -- the limit $\mu \ra 0$ does exist and
corresponds to the pole mass.

However, accounting  for the running of $\alpha_s$ makes the result
for the  pole mass  uncertain\cite{pole,bb}. Indeed, let us substitute
$\alpha_s(k^2)/k^2$ for the gluon propagator in the simplest
diagram discussed above. Then higher order terms appear in $m_Q(\mu)$:
\be
m_Q(\mu) - m_Q (\mu_0) = \frac{4\alpha_s}{3\pi} \mu \sum 2^n n!
\left(\frac{b\alpha_s}{4\pi}\right)^n,
\label{dm}
\ee
where $b$ is the one-loop $\beta$ function coefficient.
The factorial divergence of this expansion reflects an
appearance of infrared renormalons in the problem. One can try
to sum up the expansion \ref{dm} but different prescriptions
lead to different results with an uncertainty $\sim \Lambda_{QCD}$.

The relative uncertainty $\Lambda_{QCD}/m_Q$ seems to be a clear
contradiction to the statement discussed above about the absence
of $1/m_Q$ corrections to inclusive widths (see the equation
\ref{gamma} ). The parodox is simply resolved since  inclusive
 widths are defined
by short distances $\sim\, 1/m_Q$ and masses should be
correspondingly taken at deeply Eucledian distances as well.
Only in terms of these masses does the statement that there is
no $1/m_Q$ corrections make sense.

Of course, in any given order in $\alpha_s$ it is possible to
formulate results in terms of the pole mass. It means, however,
the factorial behaviour of coefficients in perturbation theory
and the renormalon uncertainty. On the other hand, the consistent
use of masses normalized at the relevant distances shows that both
the pole mass and infrared renormalons in the coefficients are
physically irrelevant.

It is just the place where the standard HQET occurs to be
inconsistent because the notion of normalization point was not
introduced there for power corrections. There was a hot discussion
in the literature (see e.g. paper\cite{neubsach}, see also
paper\cite{martsach} where the lattice is used for
nonperturbative definition of the pole mass) concerning a
consistency of HQET. I do not think that the necessity of
explicit introduction of normalization point can be avoided.

\section{Sum Rules}
In semileptonic $b \ra c$ transitions the $c$ quark can be treated
also as a heavy one. This leads to some number of sum rules
for the moments of spectral distributions such as:
\be
I_n({\vec q})=\frac{1}{2}\left(\delta_{kl} - \frac{q_k q_l}{{\vec
q}^2}\right) \sum_i \epsilon^n_i \langle B|j_k|i \rangle
\langle i|j_l|B \rangle
\ee
where $\epsilon_i = \sqrt{M_i^2 + {\vec q}^2} - \sqrt{M_0^2 +
{\vec q}^2}$ is an excitation energy of i-th state moving with
momentum $(- {\vec q)}$. The current $j_k$ here is the axial current
$j_k={\bar c}\gamma_k \gamma_5 b$. Referring to\cite{sr} for details
let us discuss few sum rules arising in slow velocity\cite{sv} (SV) limit.
\be
I_0 ({\vec q}=0)=1 - \frac{\mu_G^2}{3m_c^2} - \frac{\mu^2_\pi -
\mu^2_G}{4}\left(\frac{1}{m_c^2} + \frac{1}{m_b^2} +
\frac{2}{3m_cm_b} \right).
\label{I0}
\ee
The radiative corrections are omitted here. The unity in this sum
rule corresponds to Bjorken sum rule\cite{bjorken} at zero recoil, and the next
terms are the
$1/m_Q^2$ corrections. Numerically the term $\mu^2_G/3m_c^2$
decreases the total normalization by about 7\%.

The $1/m_c^2$ corrections are much larger for the derivative of $I_0$,
\be
m_c^2 \frac{dI_0}{d {\vec q}^2}|_{{\vec q}=0}=-\frac{1}{4}\left[
1-3\frac{\mu_G^2}{m_c^2}
 -(\mu^2_\pi -
\mu^2_G)\left(\frac{5}{2m_c^2} + \frac{1}{2m_b^2} +
\frac{1}{m_cm_b} \right)\right]\;.
\label{dI0}
\ee
Corrections now are about 50\%!

In this and in the previous sum rule terms with $\mu^2_\pi -\mu^2_G$
only enhance the effect -- this difference is positive as a
consequence of certain some rule. It gives a lower bound for
$\mu^2_\pi\,\geq \mu^2_G = 0.36\, {\rm GeV}^2$. The uncertainty
in this bound due to radiatvie corrections was analyzed\cite{sr}
and numerically is about $0.1\,{\rm GeV}^2$.

The derivative of the first moment looks as follows:
\be
m_c^2\frac{dI_1}{d{\vec q}^2}|_{{\vec q}=0}=\frac{1}{2}\left[
{\bar \Lambda} -
\frac{{\bar \Lambda}^2}{m_c} - \frac{4 \mu^2_G}{3m_c}
 -\frac{\mu^2_\pi - \mu^2_G}{2m_c}\left(\frac{7}{6}+
\frac{m_c}{3m_b}\right) \right].
\label{optical}
\ee
The first term ${\bar \Lambda}$ presents Voloshin sum
rule\cite{optical}, the next ones are the $1/m_Q$ corrections.
 The term $-4\mu^2_G/3m_c \approx -0.3\,{\rm GeV}$.

The derivative of the second moment gives the sum
rule\cite{fourth} for $\mu^2_\pi$:
\be
m_c^2 \frac{dI_2}{d {\vec q}^2}|_{{\vec q}=0}= \frac{\mu^2_\pi}{3}\;.
\label{fourth}
\ee

We see from examples above that $c$ quark is not heavy enough to guarantee
 a smallness of power corrections.

\section{Extraction of $|V_{cb}|$}
There are two sources for extraction of the value of $|V_{cb}|$:
the rate of exclusive decay $B\ra D^* l \nu$ in the zero-recoil limit
and the inclusive total width of semileptonic decay $B\ra X_c l \nu$.
The main problem is a reliable estimate of theoretical uncertainty.
Two years ago a common belief was that the exclusive decay rate gives
a more accurate extraction. Now it is known to be not the case -- the
inclusive width leads to the better accuracy.

What has been
changed? First it was a demonstration\cite{suv} of relatively large $1/m_c^2$
corrections (see the equation \ref{I0} above). The left hand side
of this equation is the sum over probabilities of transitions
with $B\ra D^*$ as a leading one. The main uncertainty comes
from the conribution of excited states into this sum what gives
additional  $1/m_c^2$ corrections. The amplitude of $B\ra D^*$
transition can be estimated as
$$
F_{B\ra D^*}\approx 0.9 \pm 0.035
$$
while in 1993 it was\cite{falkneub}  $1.00 \pm 0.04$.

Second, regarding the inclusive width it was no\-ti\-ced\cite{suv}
that its dependence on the value of $m_b$ is not so strong as it
looks like from the $m_b^5$ factor -- it depends the most
strongly on $m_b - m_c$ which is defined by $M_B - M_D$.
The uncertainty in the value of $\mu^2_\pi$ becomes important
but it can be improved when $\mu^2_\pi$ will be extracted
directly from the data. The result for $|V_{cb}|$ extracted from
 the inclusive width is (see\cite{bbb95,uraltsev}):
\be
|V_{cb}|=0.408\left[\frac{Br(B\ra X_c l
\nu)}{0.105}\right]^{1/2}
\left[\frac{1.6\,ps}{\tau_B}\right]^{1/2} (1 \pm 0.03).
\ee

\section{Semileptonic Branching Ratio}
The theoretical understanding of this branching ratio is still
lacking.  Experimentally\cite{pdg} it is $(10.43 \pm 0.24)\%$,
the theoretical situation over time is presented in
the  following table.
\begin{table}[h]
\begin{center}\caption{Semileptonic Branching Ratio (in \%)
}
\vspace{0.5cm}
\begin{tabular}{c|c|c|c} \hline\hline
&Parton & HQE~\cite{baffling} & BBBG~\cite{bbbg} \\
$\alpha_s(m_Z)$ & model~\cite{altarelli} ('91) &  ('94) & ('94) \\ \hline
& & & \\
0.110 & 13.3 & 13.2 & 12.3 \\ \hline
& & & \\
0.117 & 13.0 & 12.8 & 11.8 \\ \hline
& & & \\
0.124 & 12.5 & 12.3 & 11.3 \\ \hline\hline
\end{tabular}
\end{center}
\end{table}
The second column gives results of leading and next-to-leading log
summation of radiative corrections\cite{altarelli} for different
values of $\alpha_s$. The third column accounts for power
corrections\cite{baffling}.
The last column from the paper\cite{bbbg} which gives the lowest
 values for the branching
ratio accounts for two recent developments -- first, the
enhancement of nonleptonic $b\ra {\bar c} c d$ transition by
radiative corrections\cite{volosh,bbbg}, second, for the finite
value  of $m_c/m_b$ in the radiative loops\cite{bbbg}.

Although the theoretical branching ratio went down the problem
does not seem resolved. The point is that the
prediction\cite{bbbg} for the yield of $c$ quarks in $B$ decays is
too high, $\langle n_c \rangle =1.28 \pm 0.08$. The
experimental value is $\langle n_c \rangle =1.129 \pm 0.046$
(see the talk by Browder at this Conference).

\setcounter{secnumdepth}{0} 

\section{Acknowledgments}
I am grateful to Ikaros Bigi, Mikhail Shifman,
Nikolai Urlatsev. This talk is based on the collaboration with them.
My thanks to Larry McLerran for his help with the preparation of this
text.

\end{document}